\documentclass[12pt,aps,preprint,nofootinbib]{revtex4}
\pdfoutput=1
\usepackage{graphicx}
\usepackage{cancel}
\usepackage{amssymb}
\usepackage{amsmath,dsfont,textcomp}
\usepackage{bm}
\usepackage{color}
\usepackage{euscript,amsmath}

\newcommand{\spsi}{S_{\psi K}}

\newcommand{\afss}{a^s_{\rm fs}}
\newcommand{\asld}{a^d_{\rm SL}}

\newcommand{\GeV}{~\text{GeV}}

\def\beq{\begin{equation}}
\def\eeq{\end{equation}}
\def\bea{\begin{eqnarray}}
\def\eea{\end{eqnarray}}
\def\to{\rightarrow}

\def\hgb{horizontal gauge boson~}
\newcommand{\nn}{\nonumber}

\begin{document}
\preprint{IPMU10-0115}
\title{TeV scale horizontal gauge symmetry and its implications in $B$-physics}
\author{Seong Chan Park,$^1$ Jing Shu,$^{1,2}$ Kai Wang$^1$ and Tsutomu T. Yanagida$^1$}
\affiliation{
$^1$ Institute for the Physics and Mathematics of the Universe,\\ University of Tokyo, Kashiwa, Chiba 277-8568, JAPAN \\
$^2$ INPAC, Department of Physics, Shanghai JiaoTong University, Shanghai, 200240, CHINA
}

\begin{abstract}
We propose a gauged $U(1)_{H}$ horizontal symmetry around TeV scale that is a subgroup of a $SU(3)_{H}$ horizontal gauge symmetry broken at ${\cal O}(10^{14}) \GeV$. The breaking generates right-handed Majorana neutrino masses through a $SU(3)_H$ sextet scalar. A particular Majorana right-handed neutrino mass matrix explicitly determines the remnant $U(1)_{H}$ at low energy which only couples to $b-s$ and $\mu-\tau$ in the gauge eigenstate. The dangerous $K-\bar{K}$, $D-\bar{D}$ mixing and $B_s \rightarrow \mu^+ \mu^-$ are kept to be safe because the relevant couplings are suppressed through high powers of small mixing angles in the fermion rotation matrix. Our analysis which applies to the general case shows that the Tevatron di-muon anomaly can be explained through the $B_{s}$ and $B_{d}$ mixing while keeping all the other experimental constraints within 90 \% C. L. For the $B$ meson decay, the $B_{s}\to \mu^{\pm}\tau^{\mp}$ is the leading leptonic decay channel which is several orders of magnitude below current experimental bound. 


\end{abstract}
\maketitle

\section{Introduction}

Horizontal gauge symmetry was proposed as an extension of the SM gauge symmetries to unify all families of quarks and leptons \cite{horizontal, Wilczek:1978xi}.  
Given the three families of quarks and leptons, $SU(3)_H$ is the most natural 
choice for the horizontal gauge symmetry. Interestingly, if one assumes all the SM fermions transform under $\bf 3$ of $SU(3)_{H}$, the anomaly free condition requires three generations of right-handed neutrinos $n_R^{i=1,2,3}$ \cite{yanagidaprd} while the right-handed neutrinos also play important roles in explaining the origin of neutrino masses. Therefore, the $SU(3)_{H}$ horizontal gauge symmetry model provides a natural scheme for the seesaw mechanism \cite{seesaw} generating small masses for light neutrinos \cite{yanagidaprd}.
The Majorana neutrino mass term for the right-handed neutrinos explicitly breaks
the $SU(3)_{H}$, thus it is often believed that the horizontal gauge symmetry should be broken at a very high-energy scale $M_{R}\sim {\cal O}(10^{14})$ GeV. Then it seems impossible to test
the $SU(3)_{H}$ gauge interactions in low-energy experiments. However, it is not always the case as we will show in detail below.  Even if some subgroup of the $SU(3)_{H}$ remains unbroken, 
the right-handed neutrinos can still acquire large Majorana masses.

Since $n^{i}_{R}$ transform as ${\bf 3}$ under $SU(3)_{H}$, the Majorana neutrino mass term can  
arise from the vacuum expectation value ({\it vev}) of an $SU(3)_{H}$ sextet  $\chi_6$, 
\beq
\overline{n^{i c}_R} \langle\chi_6\rangle_{ij} n_R^{j},
\eeq
and $M_{R} = \langle \chi_6 \rangle$. The light neutrino mass is given by the seesaw mechanism as 
$m_\nu = m_{D}^{T} (M_R)^{-1} m_D$. In order to explain the the neutrino oscillation data, 
suitable $\langle \chi_6\rangle$ and $m_{D}$ are required.
For $m_{D}$ and the other SM fermion masses, there must exist octet Higgs under $SU(3)_{H}$  in order to accommodate the correct
mass hierarchy in quarks and leptons. In addition, to minimize flavor changing effects induced by the octet Higgses, we employ a scenario with additional Higgses and singlet fermions \cite{realcp}
in which the $m_{D}$ and quark mass matrices or lepton mass matrix are all independent. 
By taking a suitable gauge choice of horizontal symmetry, we always choose the $M_{R}$ to be a diagonal matrix,
\begin{eqnarray}
M_{R} = \langle\chi_6\rangle=\begin{pmatrix}
A &0&0\\
0&B&0\\
0&0&C
\end{pmatrix}
\end{eqnarray}
The $\langle \chi_6\rangle$ structure explicitly determines the symmetry breaking. 
The $SU(3)_{H}$ is completely broken in a generic vacuum with $A\neq B\neq C (\neq A)$. However, with a specific 
vacuum of $A=B=C$ for instance, the vacuum $\langle \chi_6\rangle$ is invariant under a $SO(3)$ symmetry
and the breaking is $SU(3)_{H}\to SO(3)_{H}$. 
Being symmetric second rank tensor under $SU(3)_{H}$,  the sextet $\chi_6$ transform as $\chi_6\to U^{T} \chi_6 U$ 
where $U= e^{i \epsilon_{a}T_{a}}$ and $T_{a}$ is the generator of the horizontal symmetry. A general scheme to obtain the unbroken symmetry is derived from the condition
that if $\{T, \langle\chi_6\rangle \} = 0$, $\langle \chi_6\rangle$ is invariant under transformation defined by $T$.
To illustrate the feature of our proposal, we take a vacuum as $C=-B$. This vacuum 
$\langle\chi_6\rangle=\text{diag}(A,B,-B)$ is invariant under the $SU(3)$ 
generator $\lambda_{6}$ as
\begin{eqnarray}
\label{eq:coupling}
T=\lambda_{6}=
\begin{pmatrix}
0&0&0\\
0&0&1\\
0&1&0
\end{pmatrix}~.
\label{lambda6}
\end{eqnarray}
Consequently, one can identify the unbroken $U(1)_{H}$ gauge symmetry with generator $T$
\footnote{This $U(1)$ $T$ has an unitary equivalent representation. By taking a $45{\ensuremath{^\circ}}$ rotation $R$ between $2^{nd}$ and $3^{rd}$ axes in horizontal space, the {\it vev} becomes 
\beq
R^{T}\langle \chi_6\rangle R=\left(
\begin{array}{ccc}
A&0 &0\\
0&0&B\\
0&B&0
\end{array}
\right)~. 
\eeq
The $U(1)$ then becomes 
$T^{\prime}=R^{\dagger}T R= \text{diag}(0,1,-1)$. We take, throughout this paper, the 
basis where the Majorana mass matrix is diagonal as $M_{R}=\text{diag}(A,B,-B)$ and the unbroken $U(1)_{H}$ generator is
given by T in Eq. (\ref{lambda6}). } and it
can survive to low energy, for instance ${\cal O}$(TeV) which
may lead to interesting predictions in flavor changing neutral current (FCNC) processes.
It was also observed that if there exist horizontal gauge interactions, 
CP violation can be realized with only two generations. Explicit examples of
CP violation due to $U(1)_{H}$ and $SU(2)_{H}$ was 
discussed in \cite{horizontal}.
If the above $U(1)_{H}$ is broken at the low energy,  the horizontal gauge boson exchanges can induce additional CP violations \cite{horizontal} at low energies through quark and lepton mixings.

In the last decades, huge experimental efforts had been made in improving the measurements on CP violation in the $B$ meson system. 
Very recently, the D$\cancel{\rm 0}$ Collaboration at Tevatron has reported a large charge asymmetry in like-sign di-muon  
$A^b_{s\ell}$ in both $B_s$ and $B_d$ decays with 6.1 fb$^{-1}$. 
\beq
A^b_{sl} \text{(Exp)} \equiv \frac{N^{++}-N^{--}}{N^{++}-N^{--}} = - 9.57\pm 2.51\textrm{(stat.)} \pm 1.46 \textrm{(syst.)} \times 10^{-3} \ ,
\eeq
where $N^{++} (N^{--})$ is the event number for $b\bar{b}\to \mu^{+}\mu^{+} X (\mu^{-}\mu^{-} X)$. 
Such a large di-muon charge asymmetry has a 3.2~$\sigma$ deviation from 
the SM prediction $A^{b}_{sl}(\text{SM})=(-2.3^{+0.5}_{-0.6})\times 10^{-4}$ \cite{Lenz:2006hd} and many models have been proposed to account for this anomaly \cite{xxxx,Deshpande:2010hy,Chen:2010aq,Bai:2010kf}. The CDF has also measured $A_{sl}^b =8.0 \pm 9.0 \pm 6.8 \times 10^{-3}$ \cite{CDFdimuon}, using $1.6 ~ \textrm{fb}^{-1}$ of data, which has a positive value and large uncertainties. Combining the above two results in quadrature (include the systematic uncertainty), we have
\bea
A_{sl}^b \simeq - 8.5 \pm 2.8 \times 10^{-3} \ .
\eea  

At the Tevatron both $B_d$ and $B_s$ mesons are produced, hence $A_{sl}^b$ is related to the charge asymmetries $a_{sl}^{d,s}$ in $B_d$ and $B_s$ decays by \footnote{If the semileptonic b-hadron decays do not involve CP violating phase, then the charge asymmetry is directly related to the mixing-induced CP asymmetris in $B_d$ and $B_s$ meson oscillations. $\afss = -(1.7\pm 9.1\pm1.5)\times
10^{-3}$~\cite{flavorspecific} is measured through the time dependence of  $B^0_s\to \mu^+ D_s^- X$ and its CP conjugate at D0 and $\asld=-(4.7\pm 4.6) \times 10^{-3}$~\cite{Barberio:2008fa} at the B factory. Nevertheless, we do not use those results here due to their large uncertainty. } 
\bea
A_{sl}^b = (0.506 \pm 0.043) a_{sl}^d + (0.494 \pm 0.043) a_{sl}^s \ .
\eea
New physics (NP) contributions in $B_d$ mixing are strictly constrained (we will show it more explicitly in the parameter fit later), so only large NP contributions to the $B_s$ mixing (comparing to the other meson mixings) are allowed. For the NP contribution, if the mixing in the rotation matrix between mass eigenstate and gauge eigenstate is not huge, then one would naturally expect the $U(1)_{H}$ that maximizes $b-s$ mixing as in Eq. (\ref{eq:coupling}). Indeed, for a CKM-like rotation matrix, the gauge boson coupling matrix at the tree level in the mass basis goes like 
\beq
G \sim
\left(
\begin{array}{ccc}
  \lambda^4 & \lambda^3  & -\lambda  \\
\lambda^3 & -\lambda^2 & 1 \\
-\lambda & 1 & \lambda^2
\end{array}
\right)  \ , 
\eeq
where $\lambda$ is the Wolfenstein parameter \cite{wolfenstein} around the order of the Cabibbo angle ($\lambda\simeq 0.1$). Clearly, the meson mixings between the first two generation are highly suppressed. The NP also couples to leptons. However, their contributions to the $B$ meson decay branching ratio to electron and muon are highly suppressed (although $\lambda$ should be replaced by some small mixing of the lepton rotation matrix) \footnote{Comparing to the other NP coupling matrix in the gauge eigenstate G = diag (1, 1, a) \cite{Barger:2009qs}, which has a coupling matrix \beq
G \sim
\left(
\begin{array}{ccc}
1 & a \lambda^5  & (a-2) \lambda^3  \\
a \lambda^5 &1 & (a-1)  \lambda^2 \\
(a-2) \lambda^3 & (a-1)  \lambda^2 & a
\end{array}
\right)  \ 
\eeq in the mass eigenstate, the lepton decay branching ratio, especially the one to muon is much more suppressed in our case.}. Therefore, we focus on the phenomenology in the $B$ meson mixing and decay. 

The paper is organized as follows: in section \ref{sec:model}, we propose the specific model in which we consider in the paper. In Section \ref{sec:flavor}  we show phenomenological implications of our model on flavor physics which has subsection \ref{sec:mixing} related to meson mixing and subsection \ref{sec:decay} related to meson decay. Section \ref{sec:conclusion} contains our conclusions.

\section{The Model}
\label{sec:model}

The model starts with a gauged $SU(3)_{H}$ model at extremely high energy. 
By taking all the fermions as ${\bf 3}$ under $SU(3)_{H}$. 
The particle contents under $SU(3)_{H}\times SU(2)_{L} \times U(1)_{Y}$ is
\bea
q_{L}: (3,2,\frac{1}{3}), & u_{R}: (3,1,\frac{4}{3}), & d_{R}: (3,1,-\frac{2}{3})\nonumber\\
\ell_{L}: (3,2,-1),  & e_{R}: (3,1,-2), & n_{R}: (3,1,0)
\eea 
which is exactly vectorial and the $SU(3)_{H}$ is therefore anomaly free symmetry. 
It is crucial to have right-handed neutrino triplets, $n_{R}^{ i=1,2,3}$, for anomaly cancellation \cite{yanagidaprd}. The extension 
to the Pati-Salam unification \cite{patisalam} may be straightforward. 

As we have already discussed the sextet breaking in the introduction, here we focus on the Yukawa interactions for
the other SM fermions and the Dirac neutrino mass matrix.
In conventional $SU(3)_{H}$ models, in order to break the $SU(3)_{H}$ as well as the $SU(2)_{L}\times U(1)$, one usually introduces 
one $H: (1,2,1)$, four $\Phi_{8}: (8,2,1)$ to generate all the SM fermion mass hierarchies \footnote{Another possibility is to consider bulk $SU(3)$ horizontal gauge symmetry broken at one boundary brane where all fermion masses are generated at that brane. In this case, one do not need color octet Higgses to generate SM fermion hierarchies. See Ref. \cite{Cacciapaglia:2007fw} as an example. }. However, the $(8,2,1)$ Higgs will induce large FCNC \cite{Glashow:1976nt} if the Higgs is light. To avoid the too large FCNC problem, another proposal is to introduce $(8,1,0)$ Higgs\cite{realcp}. 
\beq
\Phi^{i}_{8}: (8,1,0), H: (1,2,1)~~~~(i=u,d,e,\nu)
\eeq
In addition, to generate effective Yukawa couplings, a new set of $SU(2)_{L}$ singlet fermions is introduced
\begin{eqnarray}
U_{L}: (3,1,\frac{4}{3}), & D_{L}: (3,1,-\frac{2}{3}), E_{L}: (3,1,-2), N_{L}: (3,1,0) \nonumber\\
U_{R}: (3,1,\frac{4}{3}), & D_{R}: (3,1,-\frac{2}{3}), E_{R}: (3,1,-2), N_{R}: (3,1,0)~.
\end{eqnarray}
These singlet fermions form invariant Dirac masses and act as messengers to generate the necessary Yukawa interactions. 
We take the up-type quark mass matrix as an example. Since the octet Higgs is no longer $SU(2)_{L}$ doublet, 
$q_{L} \bar{u}_{R} \Phi$ is forbidden and the up-quark Yukawa interactions only arise as  
\beq
U_{L}\Phi_{8}^{(u)}\bar{u}_{R}+M^{U} \bar{U}_{L}U_{R}+q_{L} \bar{U}_{R} H+ \lambda_u q_{L} \bar{u}_{R} H
\eeq
where $q_{L}\bar{u}_{R}H$ is universal. 
After integrating out the heavy fermion fields $U_{L}$, $U_{R}$, the effective up-quark Yukawa coupling reduce to
\beq
 \bar{u}_{R}^i (\lambda_u \delta_{ij} + (\langle \Phi_8^{(u)} \rangle  M_U^{-1})_{ij})q_{L}^j H.
 \label{yuka}
\eeq
The same mechanism also applies to the mass generation of down type quarks, charged leptons
as well as Dirac neutrinos. By assigning the $\langle\Phi^{(i)}_{8}\rangle$ independently, 
the mixings and masses in different fermion sectors are completely independent for each other
and one can easily accommodate hierarchies and mixings in SM fermions
and the Dirac neutrinos. This also enables us to choose the Dirac neutrino mass
matrix other than nearly-diagonal structure. 

After electroweak symmetry breaking, the effective Yukawa coupling of $\Phi_8^{(u)}$ also arises as 
\beq
        \langle H\rangle  {M_U}^{-1} {\bar u}_R\Phi_8^{(u)} u_L~.
\eeq
Then, the $\Phi_{8}^{(i)}$ exchanges induce FCNC's in general. We have checked that they satisfy the strongest constraint from K-K mixing, marginally \footnote{
Since $M\simeq \langle\Phi_{8}\rangle= M_{Z^{\prime}}/g_{H} \simeq 50$~TeV, the effective coupling here is $\langle H\rangle /M$ of ${\cal O}(10^{-3})$. With additional propagator suppression due to $1/M^{2}$,
the amplitude is ${\cal O}(10^{-15})~\text{GeV}^{-2}$.}. 
However, actual effects depends on the mass spectrum of the $\Phi_{8}^{(i)}$ and hence we do not discuss them in this paper.

Another consequence is that both up and down quark mass matrices become Hermitian 
\beq
m^{\dagger}_{u} = m_{u}, m^{\dagger}_{d} = m_{d}
\eeq
Thus, the CP violation in strong interactions due to quark mass matrices,
$\text{arg}\{\det(m_{u})\det(m_{d}) \}$ is  absent at least at the tree level\cite{realcp}. 
In addition, the Hermit mass matrices also require the rotations $U_{L},U_{R}$ in the mass diagonalization $U^{\dagger}_{L} m_{u} U_{R}$ to be equal $U_{L}=U_{R}$. 
In this case, the horizontal gauge boson couples to vector currents of quarks and leptons. As a consequence, pseudo-scalar bosons like $B_{s}$ or $B_{d}$ do not decay to a pair of leptons. 
However, this is  only the result of our specific choice of mass generation model for quarks and leptons. 
In the following analyses, we assume more generic rotation matrices and $U_R$ and $U_L$ are taken independent for each other to 
estimate the prediction. 

\section{Phenomenological implications in flavor physics}
\label{sec:flavor}

The horizontal gauge interaction is real but family dependent. After the mass diagonalization, 
the other flavor violation entries as well as new CP violation can arise.
The Lagrangian of gauge interactions is
\bea
-{\cal L}_{H}& = & g_{H} \bar{q'}_L T \gamma^{\mu}{q'}_L{Z^{\prime}_{\mu}} +L\leftrightarrow R\nonumber\\
&=& g_{H}\bar{q}^i_{L} \left({V^{q}_{L}}^{\dagger}T V^{q}_{L}\right)_{ij}  \gamma^{\mu} q_{L}^j Z^{\prime}_{\mu}+L\leftrightarrow R~,
\eea
where $V^{q}_{L}$ stands for the rotation for left-handed $q$-type quarks and $T$ is the generator of $U(1)_H$ interaction given in Eq. (\ref{lambda6}).

Flavor changing interactions in the SM can only be measured via electroweak charged current interactions. Therefore,
for the SM fermion rotation matrixes, only the left-handed ones get constrained from the CKM matrix $V_L^u (V_L^d)^\dagger = V_{CKM}$
and one cannot determine even $V^{u}_{L}$ and $V^{d}_{L}$ respectively. The other rotations are completely unknown.  
For simplicity of the discussion here, we will assume that all magnitudes of the left-handed mixings are CKM-like but the complex phases are $\cal O$(1)  and unconstrained right-handed mixings have the similar structure. Therefore, we have the mixing matrix in the mass eigenstates as 
\beq
(G^{\prime})_{L/R}^{u/d} = (V^{u/d}_{L/R})^{\dagger} T ( {V^{u/d}_{L/R}}) \rightarrow G^\prime \sim
\left(
\begin{array}{ccc}
  \lambda^4 & \lambda^3  & -\lambda  \\
\lambda^3 & -\lambda^2 & 1 \\
-\lambda & 1 & \lambda^2
\end{array}
\right) 
\eeq
This $U(1)_{H}$ gauge interaction maximizes the mixing in between second and third generations.
Mixing magnitude in $B_{s}$, $B_{d}$ and $K^{0}$ is at the order of $(1:\lambda^{2}:\lambda^{6})$. 
The $D^{0}-\bar{D}^{0}$ mixing is also at the order of $\lambda^{6}$ suppression comparing with
$B_{s}$ mixing. This $U(1)_{H}$ is consistent with the phenomenological constraints among
different meson mixings. If one assume the lepton doublet and right-handed singlet rotations
are the similar to the quark sector \footnote{Within the minimal $SU(5)$ or $SO(10)$ grand unification theory (GUT),
both left-handed and the conjugate of right-handed states are embedded into the
same GUT multiplet. But in the $SU(3)_{H}$, left-handed states and right-handed states both
transform as $\bf 3$. The GUT multiplet will contain both $\bf 3$ and $\bar{\bf 3}$ and the horizontal 
gauge symmetry model is not consistent with the minimal GUT. Therefore, we will not assume any
correlation among rotations of quarks and leptons.}, one can also compute the 
leptonic decay of mesons. For instance, $B_{s}\to \mu^{+}\mu^{-}$ decay partial width
has a $\lambda^{4}$ suppression.

\subsection{Meson Mixing}
\label{sec:mixing}

At the energy scale $m_b$, the effective
Hamiltonian responsible for neutral meson mixing (and in particular
 $B_s -\bar B_s$ mixing) through the tree-level exchange of $Z^{\prime}$ is 
 \begin{eqnarray}
{\cal H}= C_{LL}^{ij}(m_b) O^{ij}_{LL} +C_{RR}^{ij} (m_b) O^{ij}_{RR} +
C_{LR}^{ij} (m_b) O_{LR}^{ij} + \widetilde{C}_{LR}^{ij} (m_b) \widetilde{O}_{LR}^{ij} ,
\label{eq:deltab2}
\end{eqnarray}
where the $\Delta F =2 $ operators are given by
\begin{eqnarray}
&&O_{LL}^{ij} =\bar q_i \gamma^\mu P_L q_j \bar q_i\gamma_\mu P_L q_j,
\;\;O_{RR}^{ij} =\bar q_i \gamma^\mu P_R q_j \bar q_i\gamma_\mu P_R
q_j,\nonumber\\
&&O_{LR}^{ij} =\bar q_i \gamma^\mu P_L q_j \bar q_i\gamma_\mu P_R q_j,
\;\;\widetilde O_{LR}^{ij} =\bar q_i P_L q_j \bar q_iP_R q_j.
\label{eq:deltab2o} \ ,
\end{eqnarray}
and the Wilson coefficients at $M_{Z^\prime}$ scale are ($\widetilde{C}_{LR}^{ij}  (M_{Z^\prime})=0$)
\beq
C_{LL}^{ij} (M_{Z^\prime}) = {g^{2}_{H} \over M^{2}_{Z^\prime}} (G_{L}^{\prime ij})^2 ~~~
C_{RR}^{ij} (M_{Z^\prime}) = {g^{2}_{H} \over M^{2}_{Z^\prime}} (G_{R}^{\prime ij})^2 ~~~
C_{LL}^{ij} (M_{Z^\prime}) = {g^{2}_{H} \over M^{2}_{Z^\prime}} (G_{L}^{\prime ij})^2
\eeq
where $g_{H}$ is the horizontal gauge coupling at $M_{Z^\prime}$ scale and $M_{Z^\prime}$
is the \hgb mass.

In order to calculate the $B$ physics observables, one has to take into account the running effects of the four operators above. The relation between these four operators at the $M_{Z^\prime}$ and $m_b$ scale is presented in Appendix A. After one obtains the Wilson coefficients at $m_b$ scale, by using the relevant hadronic matrix elements \cite{Buras:2001ra}
\bea
\langle B_q | O_{LL/RR}^{bq} | \bar{B}_q \rangle & \approx& \frac{1}{3} m_{B_q} f_{B_q}^2 B^{bq}_{LL/RR} \nonumber \\
\langle B_q | O_{LR}^{bq} | \bar{B}_q \rangle &\approx &- \frac{1}{6} m_{B_q} f_{B_q}^2 B^{bq}_{LR} \nonumber \\
\langle B_q | \widetilde{O}_{LL}^{bq} | \bar{B}_q \rangle &\approx & \frac{1}{4} m_{B_q} f_{B_q}^2 \widetilde{B}^{bq}_{LR} \eea 
Here we use $m_{B_q}^2 / (m_b + m_q)^2 \approx 1$ and assume $B^{bq}_{LR} \simeq \widetilde{B}^{bq}_{LR} \simeq B^{bq}_{RR} =  B^{bq}_{LL} \equiv B_{B_q}$. Then we obtain $M_{12}^q$  
\begin{eqnarray}
\label{m12m}
&&M^{q}_{12} \equiv \langle B_q |  \mathcal{H}  | \bar{B}_q \rangle = -
{1\over 3} f_{B_q}^2 m_{B_q} B_{B_q} \left [ C^{bq}_{LL}(\mu) +
C^{bq}_{RR}(\mu) - \frac{ C^{bq}_{LR}(\mu) }{2} +
\frac{3\tilde C^{bq}_{LR}(\mu)}{4} \right ] \ .
\end{eqnarray}

From the discussion above, the flavor off diagonal coupling between the horizontal gauge boson $Z^{\prime}$ and the first two generation quarks are highly suppressed (For the CKM like rotation matrix, it is at least $\lambda^3$ suppressed), so we will neglect the new physics contributions to the $K - \bar{K}$ 
\footnote{The constraint from the CP violation in the $K-\bar{K}$ mixing \cite{Bona:2007vi} is, in fact, marginal for $\lambda \simeq 0.1$, $g_{H}\simeq 0.02$ and 
$M_{Z^{\prime}}\simeq 1$~TeV \cite{horizontal}. } and $D - \bar{D}$ mixing. The $Z^{\prime}$-$b$-$s$ and $Z^{\prime}$-$b$-$d$ couplings, on the other hand, is either unsuppressed or $\lambda$ suppressed, hence we expect large new physics contributions to modify the magnitudes and the phases of $M_{12}^{d/s}$, where $M_{12}^{d/s}$ are off-diagonal mixing matrix elements in Eq.\ref{m12m}. We can parametrize such effects by \footnote{The CP violation phase is defined as $\phi \equiv \textrm{Arg}(- M_{12} / \Gamma_{12} )$ so we choose $\Gamma_q$ to be real here.} 
\bea
M_{12}^{d/s} \equiv (M_{12}^{d/s})^{\textrm{SM}}  \Delta_{d/s} ~~~~~~ \Delta_s \equiv | \Delta_{d/s} | e^{i \phi_{d,s}^\Delta} .
\eea

\begin{figure}
  \includegraphics[width=8cm]{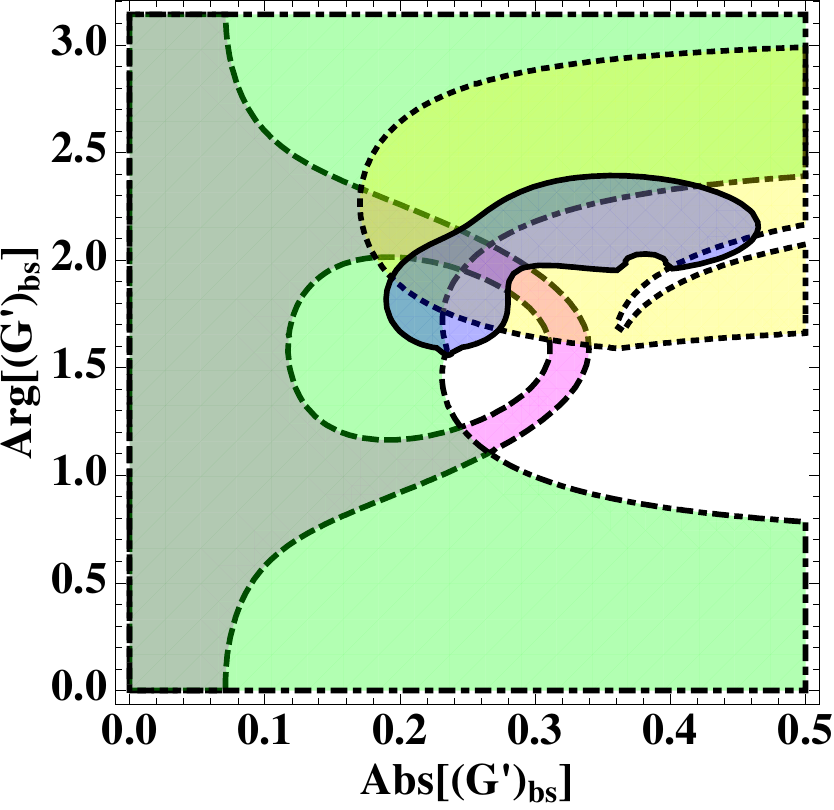}
  \includegraphics[width=8cm]{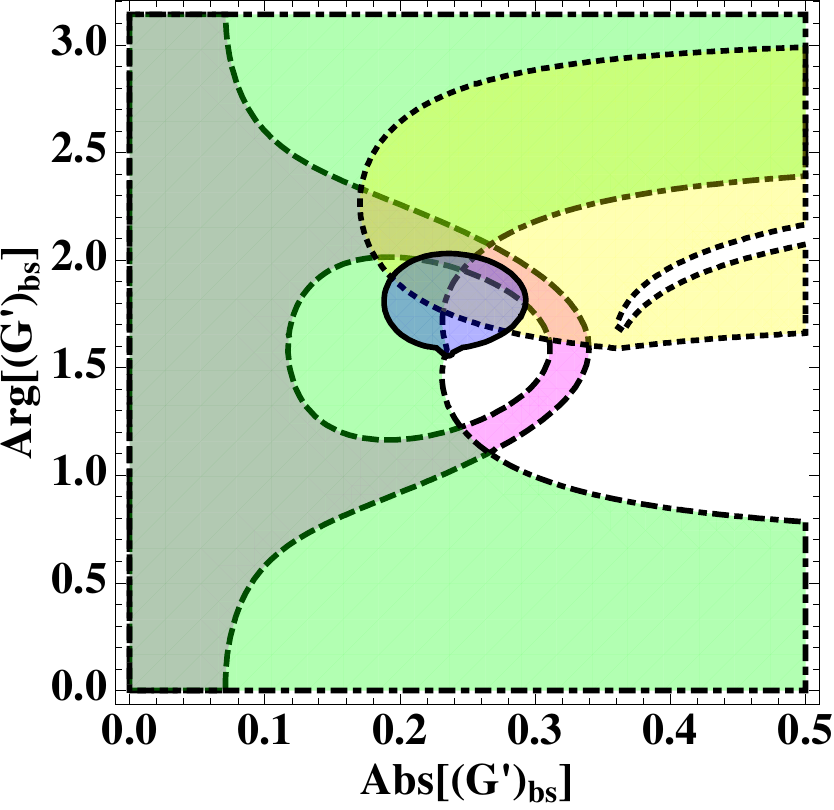} \\
    \includegraphics[width=8cm]{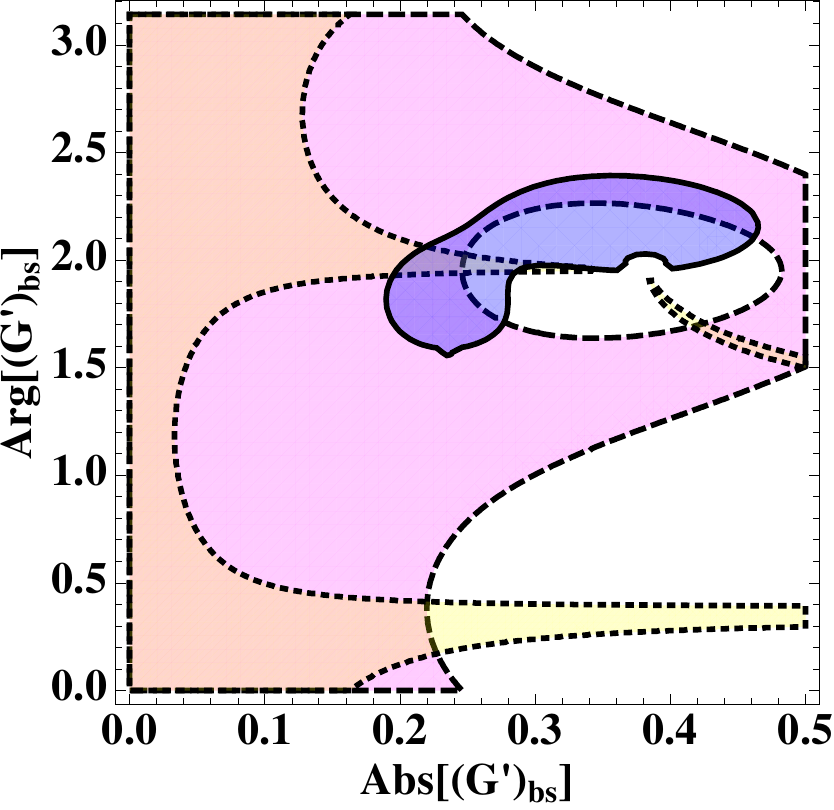} 
      \includegraphics[width=8cm]{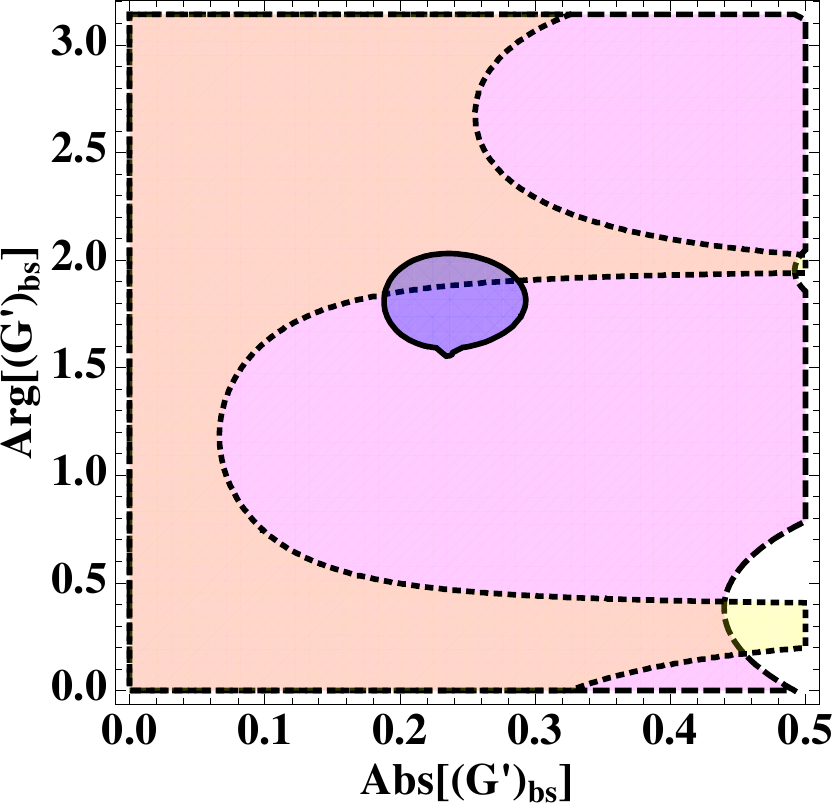} 
  \caption{The Abs$(G^\prime_{bs})$ versus Arg$(G^\prime_{bs})$ region plot for $M_{Z^\prime} = 1$ TeV and $g_H=0.02$. In the left panel, we choose Abs$(G^\prime_{bd})$/Abs$(G^\prime_{bs}) = 0.1$ and Arg$(G^\prime_{bd})$ = Arg$(G^\prime_{bs})$ while in the right panel, we choose Abs$(G^\prime_{bd})$/Abs$(G^\prime_{bs}) =0.05$ and Arg$(G^\prime_{bd})$ = Arg$(G^\prime_{bs})$. The magenta, green, and yellow region with dashed, dot-dashed and dotted boundary stands for the allowed parameter space at  90\% C. L. for $\Delta m_{{s/d}}$, $\Delta \Gamma_{s}$ and $\spsi$/$\beta_s$.}
  \label{fig:mtmb}
\end{figure}

\begin{figure}
  \includegraphics[width=8cm]{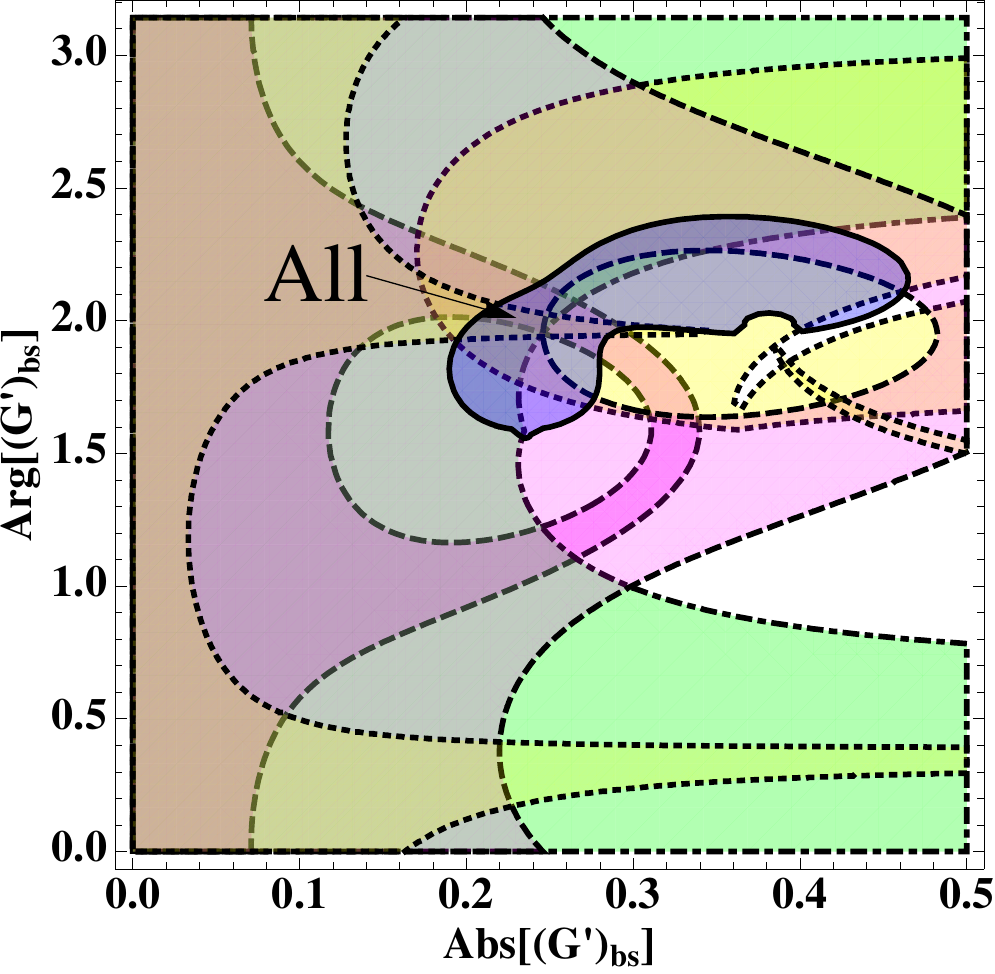}
    \includegraphics[width=8cm]{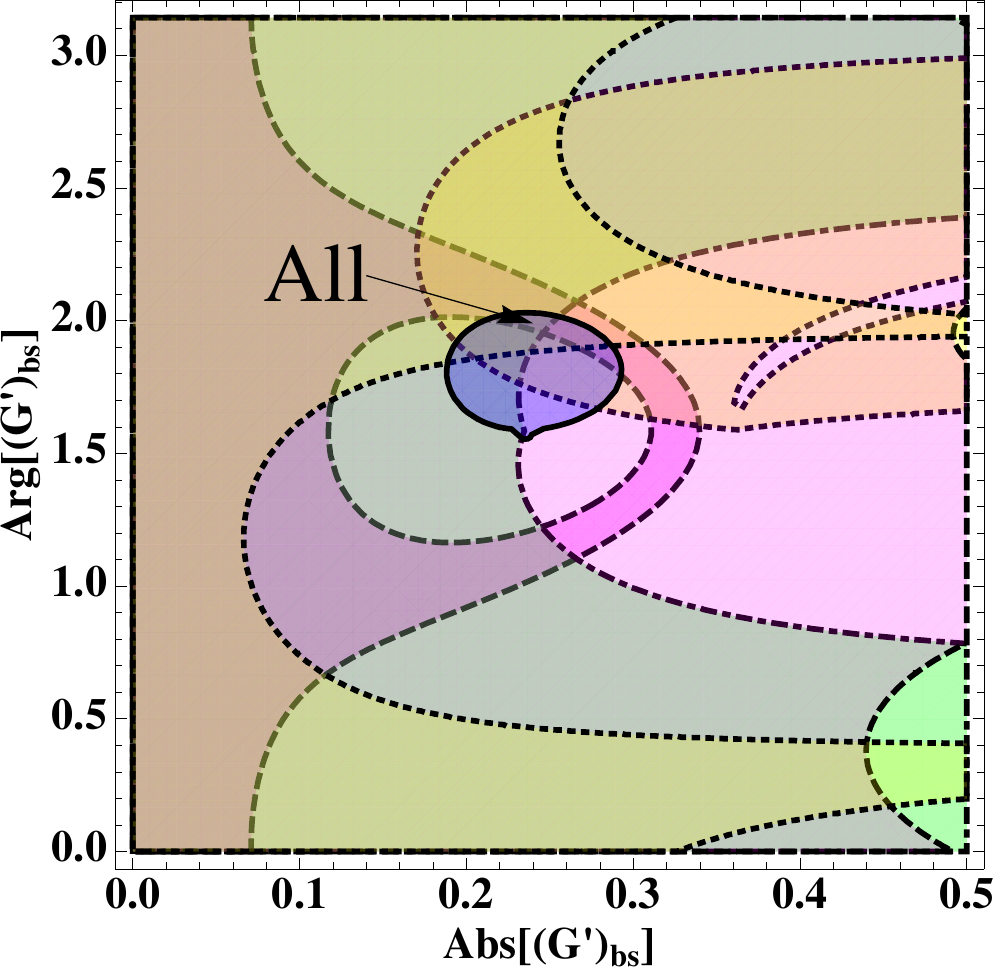} 
  \caption{The similar plots as FIG. \ref{fig:mtmb} but combine both the $B_d$ and $B_s$ experimental constraints. The small overlapped region labeled with ``All" is the parameter space that fits for all experimental constraints at 90\% C.L. 
  }
  \label{fig:mtmball}
\end{figure}

The experimental measured observables are summarized as follows, $\Delta m_{d/s}$ and $\Delta \Gamma_{d/s}$ measures the mass and decay width difference between the heavy and light mass eigenstates of the $B_{d/s}$ mesons \footnote{The experimental uncertainty in the measurements $\Delta \Gamma_{d} = \textrm{sign}(\textrm{Re} \lambda_{C P}) \Delta \Gamma_d / \Gamma_d = 0.009 \pm 0.037.$ is too big for us to consider it.}. $a_{SL}^{d/s}$ is the charge asymmetry in semileptonic $B_{d/s}$ decays. $\beta_d$ or $\beta_s$ measure the time-dependent CP violating phases in the hadronic $B$ decay channel $B_d \rightarrow J / \psi~  K_S$ or  $B_s \rightarrow J / \psi ~ \phi$. They are shifted by the CPV phases in $B_d$ or $B_s$ mixing.  
\bea
\label{eq:parametrization}
\Delta m_{d/s} &=& \Delta m_{d/s}^{\rm SM}\,
  \big| \Delta_{d/s} \big| \,, \nn\\
\Delta\Gamma_s &=& \Delta\Gamma_s^{\rm SM}\,
  \cos (\phi_s^{\rm SM} + \phi_s^\Delta ) \,,\nn\\
a^{d/s}_{\rm SL} &=&   \frac{ \Delta\Gamma_{d/s}^{\rm SM} }  { \Delta m_{d/s}^{\rm SM} } \frac{\sin (\phi_{d/s}^{\rm SM} + \phi_{d/s}^\Delta ) }{ | \Delta_{d/s} | } \,,\nn\\
 \spsi &=& \sin \big(2\beta_d +  \phi_d^\Delta \big) \,, \nn\\
2 \beta_s^{\rm Exp} &=&  2\beta_s - \phi_s^\Delta  \,.
\eea

\begin{table}[hptb]
\caption{The theoretical input parameters  \cite{Lenz:2006hd}.
 } \label{tab:inputs}
\begin{ruledtabular}
\begin{tabular}{cccc}
   $\beta_d$ & $\beta_s$ & $m_{B_d}$ & $m_{B_s}$ 
 \\ \hline
  $0.38 \pm 0.01$ \cite{Bona:2009tn} & $0.018 \pm 0.001$ & 5.28 GeV & 5.37 GeV  \\ \hline\hline
 $f_{B_d} \sqrt{\hat B}_d$ & $f_{B_s} \sqrt{\hat{B_s}}$ & $f_{B_d}$ & $f_{B_s}$
   \\ \hline
 $(216\pm 15)$ MeV & $(275 \pm 13)$ MeV & $192.8 \pm 9.9 $ MeV & $238.8 \pm 9.5$ MeV   \\ \hline\hline
 $\phi^{\rm SM}_d$& $\phi^{\rm SM}_s$  & $(\Delta m_{B_d})^{\rm SM}$ & $(\Delta m_{B_s})^{\rm SM}$      \\ \hline
 $-0.091^{+0.026}_{-0.038}$ & $(4.2 \pm 1.4) \times 10^{-3}$ & $0.53 \pm 0.12$ ps$^{-1}$ &  $19.30 \pm 2.2$ ps$^{-1}$   \\ \hline\hline
$(\Delta \Gamma_d)^{\rm Exp}$ & $(\Delta\Gamma_s)^{\rm Exp}$        \\ \hline
$(2.67^{+0.58}_{-0.65})\times 10^{-3} \ {\rm ps^{-1} }$ & $0.098 \pm 0.024 \ {\rm ps^{-1} }$     \\ \hline
\end{tabular}
\end{ruledtabular}
\end{table}

The theoretical inputs are listed in the Table \ref{tab:inputs}. All the decay constants and bag parameters are used from Ref. \cite{Laiho:2009eu}. Notice that we use the calculations for $\Delta m_{B_s} = 2 |M_{12}^s|^{SM}$, $\Delta m_{B_s} = 2 |M_{12}^s|^{SM}$ and $\phi^s$ in Ref. \cite{Bauer:2010dg} which uses the more recent decay constants and bag parameters with much smaller uncertainties. All of the rest SM inputs are either from \cite{Lenz:2006hd} or \cite{PDG08}

All experimental measurements which are used to compare with our model outputs are listed in Table \ref{tab:experiments}. For the like-sign dimuon charge asymmetry, we use $A_{SL}^b \simeq-(8.5 \pm 2.8) \times 10^{-3}$ which combine the D$\cancel{\rm 0}$ measurements with the CDF measurements. 
For $\beta_s^{\rm{Exp}}$ and $\Gamma_s$ measured by both CDF and D$\cancel{\rm 0}$ \cite{D0betas, CDFbetas, Tevbetas, CDFbetasnew}, we use the combined results with each measurements in 2.8 fb$^{-1}$ \cite{Tevbetas} where do not add the most recent CDF results \cite{CDFbetasnew} since it is difficult for us to extrapolated their contributions between $2.8 -5.2$ fb$^{-1}$. Nevertheless, we find that our conclusion in the paper does not change if we use the combined results in Ref. \cite{Bai:2010kf} \footnote{For the $\beta_s^{\rm{Exp}}$ at 90$\%$ C. L., we find $\beta_s^{\rm{Exp}} \subset (0.13, 0.74)$ from Fig. 7 in Ref. \cite{Tevbetas} since 1 - CL is almost linear to Log(L) where $L$ is the likelihood ratio. }. 

The experimental constraints on the parameter space of our model are presented in Fig \ref{fig:mtmb} and \ref{fig:mtmball}. The parameter Abs$(G'_{bs})$ and Arg$(G'_{bs})$ are quite similar as the parameter $h_s^2$ and $\sigma_s$ in Ref. \cite{fit} except for a overall factor with very small phase related to the  $(M_{12}^s)^{\rm SM}$, therefore our results here can be used as the model independent analysis after some re-parametrization. For illustration, we choose parameter for $B_d$ couplings Abs$(G^\prime_{bd})$ = Abs$(G^\prime_{bs})$/10, Arg$(G^\prime_{bd})$ = Arg$(G^\prime_{bs})$ in which there are sizable  contribution to $A_{sl}^b$ from $B_d$ and Abs$(G^\prime_{bd})$ = Abs$(G^\prime_{bs})$/20, Arg$(G^\prime_{bd})$ = Arg$(G^\prime_{bs})$ in which the $a_{sl}^d$ contribution to $A_{sl}^b$ is negligible. In contrast to the paper \cite{Deshpande:2010hy}, it is clearly that from the upper right plot in FIG \ref{fig:mtmb}, there is no region allowed by all experimental constraints within the 1 $\sigma$ \footnote{We notice that in \cite{Deshpande:2010hy}, they use SM prediction directly from \cite{Lenz:2006hd} which use the old decay constants and bag parameters. Nevertheless, we find that our conclusion still holds because the parameter space mentioned in \cite{Deshpande:2010hy} are not allowed by $(\Delta m_{B_s})^{\rm Exp}$ and $(\Delta \Gamma_s)^{\rm Exp}$. In the other paper which contains the model independent fit \cite{Chen:2010aq}, they simply neglect all the large uncertainties from the SM prediction which lead to the wrong upper bound of $A_{sl}^b$.}. The best fitted region for the phase Arg$(G'_{bs}) \subset (\pi/2, 3 \pi /4)$ are quite consistent with the one found in Ref. \cite{fit}. However, since the Ref. \cite{fit} essentially marginalize over $\Gamma_q^{12}$ in the range $0-0.25$ ps$^{-1}$ and use the best fit points in which $\Delta \Gamma_s$ is about 2.5 times larger than the prediction, the goodness of fit in our result here is reduced significantly comparing to the one in Ref. \cite{fit}.  

\begin{table}[hptb]
\caption{The experimental data.
 } 
 \label{tab:experiments}
\begin{ruledtabular}
\begin{tabular}{ccccc}
 $S^{\rm Exp}_{J/\Psi K_S}$& $ (\beta_s)^{\rm Exp}$  & $(\Delta m_{B_d})^{\rm Exp}$ & $(\Delta m_{B_s})^{\rm Exp}$  & $(\Delta \Gamma_s)^{\rm Exp}$    \\ \hline
 $0.655 \pm 0.024$ \cite{Barberio:2008fa} & $0.44^{+0.17}_{-0.18}$ & $0.507 \pm 0.005$ ps$^{-1}$ & $17.77 \pm 0.12$ ps$^{-1}$ & $0.154^{+0.054}_{-0.07}$ ps$^{-1}$ \\ \hline\hline
 $A_{sl}^b$ &  ${\cal B}^{\rm Exp}(B_s\to e^+ e^{-})$  &  ${\cal B}^{\rm Exp}(B_s\to \mu^+ \mu^{-})$  & ${\cal B}^{\rm Exp}(B_s\to e^\pm \mu^{\mp})$ & ${\cal B}^{\rm Exp}(B^0 \to \mu^\pm \tau^{\mp})$  \\ \hline
 $- (8.5 \pm 2.8) \times 10^{-3}$ ps$^{-1}$ & $<5.4 \times 10^{-5}$   &  $< 4.7 \times 10^{-8}$ & $<6.1 \times 10^{-6}$ & $<3.8 \times 10^{-5}$ \\ \hline
\end{tabular}
\end{ruledtabular}
\end{table}

\subsection{Meson Decays}
\label{sec:decay}

The new horizontal gauge boson can also mediate meson hadronic decay or leptonic decays
at tree level. In this session, we choose to discuss the two leading processes, 
$b\to s c\bar{c}$ and $b\to s \mu^{\pm}\tau^{\mp}$ respectively
to illustrate how meson decays constrain the horizontal gauge interaction. 
The other transitions are always with additional factors of power of $\lambda$ suppressions.  

The effective $\Delta F =1$ Hamiltonian responsible for neutral meson decay is 
\beq
{\cal H}^{\Delta F=1}_{\rm eff}=C_{3} Q_{3}+C_{5} Q_{5}+\tilde{C}_{3} \tilde{Q}_{3}+\tilde{C}_{5} \tilde{Q}_{5}
\eeq
and the leptonic decay like $B^{0}_{d,s}\to \ell^{+}\ell^{-}$, 
\beq
{\cal H}^{\ell}_{\rm eff}=C_{9} Q_{9}+C_{10} Q_{10}+\tilde{C}_{9} \tilde{Q}_{9}+\tilde{C}_{10} \tilde{Q}_{10}
\eeq
where $i,j=e,\mu,\tau$ as lepton flavor indices. 

In the case of $b\to s c\bar{c}$ transition, SM contribution at tree level is induced via weak charged current
with a CKM factor $V_{bc} V^{*}_{sc}\sim \lambda^{2}$. Reading from the effective couplings, 
the horizontal gauge boson mediated $b\to s c\bar{c}$
has a factor of $ G_{bs}G_{cc}\sim  \lambda^{2}$. The SM and horizontal gauge interaction
contributions are at the same order of $\lambda$ and the one can simply compare
their couplings and gauge boson masses to estimate the ratio. 
As we discuss in previous section, the $U(1)_{H}$ is broken at $\cal O$(TeV) which results in a
suppression due to $(g_{H}/g)^{4} (m^{4}_{W}/M^{4}_{Z^{\prime}})\sim 10^{-8}$. Consequently, the contribution to $b\to s c\bar{c}$
from new horizontal gauge boson is completely negligible. In the case of neutral meson mixing, SM leading 
contribution is from box-diagram while the horizontal gauge boson contribution is at tree level. Therefore,
even if the new horizontal gauge boson is of order TeV, it is still possible to change the $\Delta M$ significantly.  
For decay process, if there exists SM tree level, the above argument then always applies. 
We won't discuss any constraint from such decays. 

Since the horizontal gauge boson also couples to the leptons, there is again tree level
contribution to the meson leptonic decay. 
Within the framework of SM, $B_{s}$ pure leptonic decays are realized via 
the electroweak penguin diagrams with $Z/\gamma^{*}\to \ell^{+}\ell^{-}$ and therefore, there is no lepton
flavor violation at all. The constraints on leptonic decay are mostly on
leptons directly decaying from $B$ meson. If there exists a $\tau$ in the final state,
$\tau$ decay will complicate the search due to the $D^{\pm}$ decays.
Therefore, the leading constraints are 
\begin{eqnarray}
\text{Br}(B_{s}\to \mu^{+}\mu^{-}) &< &4.7 \times 10^{-8} 	\nonumber\\
\text{Br}(B_{s}\to e^{+}e^{-}) &< & 5.4\times 10^{-5}\nonumber\\
\text{Br}(B_{s}\to e^{\pm}\mu^{\mp}) &<&  6.1 \times 10^{-6} 
\end{eqnarray}	

In our model, the horizontal gauge boson has maximal coupling to $b,s$ and $\mu,\tau$ and
the leading leptonic decay constraint is from $B_{s}\to \mu^{\pm}\tau^{\mp}$. 
But as we mentioned, $B_{s}\to \mu^{\pm}\tau^{\mp}$ does not exist in
SM physics at leading order and it is only from the new physics contribution.
Using$\langle 0\mid \bar{s}\gamma^{\mu}\gamma_{5} b \mid \bar{B}^{0}\rangle = \imath f_{B} p^{\mu}_{B}$, one can
compute the decay BR as
\beq
\text{Br}(B_{s}\to \mu^{\pm}\tau^{\mp})= \tau_{B}\Gamma(B^{0}\to \mu^{\pm}\tau^{\mp})=f^{2}_{B_{s}}\tau_{B} {m^{3}_{B}\over 64\pi}{\left( 1-{m^{2}_{\tau}\over m^{2}_{B}}\right)}^{3}{\left({m_{\tau}\over m_{B}}\right)}^{2} {C^{\mu\tau}_{9,10}}^{2}
\eeq
where the Wilson coefficients $C_{9,10} = g^{2}_{H}(m_{b})/m^{2}_{Z^{\prime}}$.
To estimate the decay BR, we take $M_{Z^{\prime}}\simeq 10^{3}$~GeV, $g_{H}\simeq 0.02$ and substitute 
$m_{b}=4.7~$GeV, $m_{\tau}=1.777~$GeV, $f_{B_{s}}=230~$MeV, $m_{B}=5.3~$GeV, $\tau_{B}=1.6$~ps,
we obtain
\beq
 \text{Br}(B_{s}\to \mu^{\pm}\tau^{\mp})\simeq 1.2\times 10^{-9} .
\eeq\footnote{The result is based on assumption from left-handed chiral interaction only which is sufficient to estimate the maximal value of leptonic decay BR.} 
The only relevant search is from CLEO as of $B_{d}\to \mu^{\pm}\tau^{\mp}$ \cite{mutau}, 
\beq
\text{Br}(B_{d}\to \mu^{\pm}\tau^{\mp})< 3.8\times 10^{-5} \ .
\eeq
Due to horizontal gauge boson given, the $B_{d}$ decay partial width has an additional $\lambda^{2}$ suppression
so the bound is prediction well below the experimental bound.
One can also estimate the $B_{s}\to \mu^{+}\mu^{-}$ using the above result. The $B_{s}\to \mu^{+}\mu^{-}$ has a factor of $\lambda^{4}$ suppression 
then the prediction is about two orders lower than the current experimental bound. 

The other possible rare decay which can be induced by the horizontal gauge boson is the 
FCNC decay in top quark, for instance, $t\to c/u +\mu^{\pm}\tau^{\mp}$. However,
given the large $M_{Z^{\prime}}$, the three body decay is highly suppressed. 
\beq
\Gamma(t\to c/u+\mu^{\pm}\tau^{\mp})= \frac{G^{2}_{F}m^{5}_{t}}{192\pi^{3}}\left({m_{W}\over M_{Z^{\prime}}}\right)^{4}\left({g_{H}\over g}\right)^{4}\sim 1.7\times 10^{-13}
\eeq
Even at the top factory like Large Hadron Collider, it is impossible to observe such rare decay event.

\section{Conclusion}
\label{sec:conclusion}

The dimuon asymmetry reported by D$\cancel{\rm 0}$ Collaboration is much larger than the SM prediction which suggests new sources for CP violation. In this paper, we propose the possibility to explain such an anomaly through a tree level exchange of a gauged $U(1)_{H}$ horizontal symmetry in $B$ meson mixing. The $U(1)_{H}$ horizontal symmetry is a remnant symmetry of $SU(3)_H$ broken at $M_R \sim \mathcal{O} (10^{14})$ GeV through a sextet scalar which gives the neutrino mass. Such a $U(1)_H$ gauge boson only couples to $b-s$ and $\mu-\tau$ in the gauge eigenstate which suppresses all other dangerous meson mixings and $B$ meson decays after the flavor rotation. For a general flavor rotation matrix we consider 
there is a parameter region around the phase Arg$(G'_{bs}) \subset (\pi/2, 3 \pi /4)$ which fits the data at 90\% C. L. For the $B$ decay, the dominate enhanced channel is the $B_s \rightarrow \mu^\pm \tau^\mp$. Nevertheless, such a enhanced decay channel is still one order of magnitude smaller than the current experimental bound.

\section{Acknowledgements}
We would like to thank Tao Liu, Hitoshi Murayama, Bai Yang and Guohuai Zhu for useful discussions. The work is partially supported by the World Premier International Research Center Initiative (WPI initiative) MEXT, Japan. S.C.P., J.S. and K.W. are also supported by the Grant-in-Aid for scientific research (Young Scientists (B) 21740172), (Young Scientists (B) 21740169) and (Young Scientists (B) 22740143) from JSPS, respectively. J.S. would like to acknowledge the hospitality of Institute of Nuclear, Particle, Astronomy and Cosmology (INPAC), Shanghai Jiao Tong University while the work was initiated.

\section*{Appendix A: RG Running of the $\Delta B = 2$ Operators}
\label{sec:app}

In the previous section, the Wilson coefficients are given at the $M_{S}$
scale while to calculate the physics processes involving low energy mesons, 
one will need to calculate relevant Wilson coefficients at the low energy scale. The running contains two steps, the first step is from $M_{Z^\prime}$ scale to $m_t$ where six flavors contribute to the running of $\alpha_s$, the second step is from $m_t$ to $m_b$, where only five flavors contribute.    
We summarize the running effects of the relevant $\Delta F=2$ operators 
below \cite{Buras:2001ra}. 

The operators belonging to the LL/RR, LR sectors read
\bea
Q_{LL} &=& (\bar{s}^\alpha \gamma_\mu P_L b^\alpha) (\bar{s}^\beta \gamma^\mu P_L b^\beta), \nonumber \\
Q_{RR} &=& (\bar{s}^\alpha \gamma_\mu P_L b^\alpha) (\bar{s}^\beta \gamma^\mu P_L b^\beta), \nonumber \\
Q_{LR} &=& (\bar{s}^\alpha \gamma_\mu P_L b^\alpha) (\bar{s}^\beta \gamma^\mu P_R b^\beta). \nonumber \\
\tilde{Q}_{LR} &=& (\bar{s}^\alpha P_L b^\alpha) (\bar{s}^\beta P_R b^\beta).
\eea
\bea
\label{eq:running}
C_{LL} (\mu_b) &=& [\eta (\mu_b)] C_{LL} (\mu_t), \nonumber \\
C_{RR} (\mu_b) &=& [\eta (\mu_b)] C_{RR} (\mu_t), \nonumber \\
\begin{pmatrix}
C_{LR} (\mu_b)\\
\tilde{C}_{LR} (\mu_b)
\end{pmatrix} &=&
\begin{pmatrix}
[\eta_{11} (\mu_b)] &[\eta_{12} (\mu_b)] \\
[\eta_{21} (\mu_b)] &[\eta_{22} (\mu_b)] 
\end{pmatrix}.\begin{pmatrix}
C_{LR} (\mu_t)\\
\tilde{C}_{LR} (\mu_t)
\end{pmatrix} 
\eea

The variables $\eta_{i}$ is defined as ratio between strong coupling
constant at different scales. Given the $U(1)_{H}$ gauge boson
is around 10 TeV, we have two different running of $\alpha_{s}$
taking into account the threshold correction due to top quark. For the running between the $B$ physics scale and top quark, we have 
\beq
\eta_{5}\equiv \frac{\alpha^{(5)}_{s}(\mu_{t})}{\alpha^{(5)}_{s}(\mu_{b})} 
\eeq
For the explicit form of $\eta$, we list both the expression at the LO (with subscript (0)) and NLO (with subscript (1))
\bea
\eta^{(0)} (\mu_b) &=& \eta_{5}^{6/23}\ , \nonumber \\
\eta^{(1)} (\mu_b) &=& 1.6273(1-\eta_5)\eta_{5}^{6/23} \ , \nonumber \\
\eta^{(0)}_{11} (\mu_b) &=& \eta_{5}^{3/23} \ , \nonumber \\
\eta^{(0)}_{12} (\mu_b) &=& 0 \ , \nonumber \\
\eta^{(0)}_{21} (\mu_b) &=& \frac{2}{3} (\eta_{5}^{3/23} - \eta_5^{-24/23}) \ , \nonumber \\
\eta^{(0)}_{22} (\mu_b) &=& \eta_5^{-24/23} \ , \nonumber \\
\eta^{(1)}_{11} (\mu_b) &=& 0.9250 \eta_5^{-24/23} + \eta_{5}^{3/23} (-2.0994 + 1.1744 \eta_5) \ ,\nonumber \\
\eta^{(1)}_{12} (\mu_b) &=& 1.3875 (\eta_5^{26/23} - \eta_5 ^{-24/23}) \ , \nonumber \\
\eta^{(1)}_{21} (\mu_b) &=& (-11.7329 + 0.7829 \eta_5) \eta_{5}^{3/23} - \eta_5^{-24/23} (-5.3048 + 16.2548 \eta_5) \ , \nonumber \\
\eta^{(1)}_{22} (\mu_b) &=& (7.9572 - 8.8822 \eta_5) \eta_5^{-24/23} + 0.9250 \eta_5 ^{26/23} \ .
\eea 

Similarly, for the running between the top quark mass scale and the horizontal gauge boson scale, one can replace all the $\mu_b$, $\mu_t$ by $\mu_t$, $\mu_{M_{Z^\prime}}$ in Eq. (\ref{eq:running}), where all the $\eta$s are
\bea
\eta^{(0)} (\mu_b) &=& \eta_{6}^{6/21}\ , \nonumber \\
\eta^{(1)} (\mu_b) &=& 1.3707(1-\eta_6)\eta_{6}^{6/21} \ , \nonumber \\
\eta^{(0)}_{11} (\mu_b) &=& \eta_{6}^{3/21} \ , \nonumber \\
\eta^{(0)}_{12} (\mu_b) &=& 0 \ , \nonumber \\
\eta^{(0)}_{21} (\mu_b) &=& \frac{2}{3} (\eta_{6}^{3/21} - \eta_6^{-24/21}) \ , \nonumber \\
\eta^{(0)}_{22} (\mu_b) &=& \eta_6^{-24/21} \ , \nonumber \\
\eta^{(1)}_{11} (\mu_b) &=& 0.9219 \eta^{-24/21} + \eta_{6}^{3/21} (-2.2194 + 1.2975 \eta_6) \ ,\nonumber \\
\eta^{(1)}_{12} (\mu_b) &=& 1.3828 (\eta_6^{24/21} - \eta_6 ^{-24/21}) \ , \nonumber \\
\eta^{(1)}_{21} (\mu_b) &=& (-10.1463 + 0.8650 \eta_6) \eta_{6}^{3/21} + \eta_6^{-24/21} (-6.4603 + 15.7415 \eta_6) \ , \nonumber \\
\eta^{(1)}_{22} (\mu_b) &=& (9.6904 - 10.6122 \eta_6) \eta_6^{-24/21} + 0.9219 \eta_6 ^{24/21} \ ,
\eea 
and 
\beq
\eta_{6}\equiv \frac{\alpha^{(6)}_{s}(\mu_{h})}{\alpha^{(6)}_{s}(\mu_{t})}  \ .
\eeq

\end{document}